\documentclass[a4paper,12pt]{article}

\usepackage{amsmath}
\usepackage{bm}
\usepackage[dvips]{geometry}
\usepackage[T1]{fontenc}
\usepackage{graphicx}
\usepackage[amssymb]{SIunits}
\usepackage{url}
\usepackage{times}
\usepackage{setspace}
\usepackage{dotlessi}
\usepackage{hyperref}
\usepackage{algorithm}
\usepackage{algorithmic}
\usepackage{amssymb}

\newcommand{\J}{\mathrm{j}} 
\newcommand{\D}{\mathrm{d}} 
\newcommand{\E}{\mathrm{e}} 
\newcommand{\imag}{\mathrm{Im}}
\newcommand{\real}{\mathrm{Re}}
\newcommand{\cI}{\mathcal{I}} 
\newcommand{\nhat}{\hat{\mathbf{n}}}
\newcommand{\ihat}{\hat{\mathbf{\dotlessi}}}
\newcommand{\jhat}{\hat{\mathbf{\dotlessj}}}
\newcommand{\khat}{\hat{\mathbf{k}}}
\newcommand{\Pbar}{\bar{P}}
\newcommand{\rhat}{\hat{\mathbf{r}}}
\newcommand{\Ynmi}[1]{Y_{n,m}^{(#1)}}
\newcommand{\Qmi}[1]{Q_{m}^{(#1)}}
\newcommand{\dQmi}[1]{\dot{Q}_{m}^{(#1)}}
\newcommand{\ddQmi}[1]{\ddot{Q}_{m}^{(#1)}}
\newcommand{\Nmi}[1]{N_{m}^{(#1)}}
\newcommand{\dNmi}[1]{\dot{N}_{m}^{(#1)}}

\graphicspath{{./Figures/}}

\begin{document}


\title{Translation of transient acoustic fields}

\author{M. J. Carley}


\maketitle

\begin{abstract}
  A method is presented for the translation of acoustic field data
  from a source to a target region. Field data are represented as
  spherical harmonic expansions on spheres surrounding the source and
  target regions respectively and expansions are translated using a
  ``point and shoot'' method using the Kirchhoff--Helmholtz integral
  to carry out an axial translation from one sphere to the other. The
  principal motivation for the method is its use in a time-domain Fast
  Multipole Method, and test cases reflective of this application are
  presented. The method converges to six digits for appropriate values
  of parameters and for the values of $N$ considered here
  computational effort scales approximately as $N^{2}$ where $N$ is
  the order of spherical harmonic expansion for the field data. The
  method is causal and thus avoids artifacts generated in methods
  which are not based on intrinsically causal formulations.
\end{abstract}

\section{Introduction}
\label{sec:introduction}

The question of how a potential field in one region may be determined
from knowledge of the field in another region is a problem which
appears in a number of
contexts~\cite{benedict-field-lau13,field-lau15,
  martin16,martin16a,greengard-hagstrom-jiang14}. In this paper, we
are interested in determining the field inside a region bounded by a
spherical surface using data on another spherical surface which
encloses time-varying acoustic sources. The principal motivation for
the work is the development of a translation method for use in a
time-domain Fast Multipole Method (FMM). A fundamental operation in
the FMM is the translation of information between source and field
regions, and there has been considerable work on how best to implement
the operation in static and time-harmonic problems (see chapter~7 of
reference\cite{gumerov-duraiswami04}). In transient calculations, the
translation operation is a topic of active
research~\cite{takahashi-tanigawa-miyazawa22}, because it limits the
overall performance of time-domain scattering codes.

In this paper, we address the problem of translating acoustic field
data from a source to a target region, using a spherical harmonic
representation of the data. The basis of the method is the
Kirchhoff--Helmholtz integral, which has been used in previous
work\cite{carley25} to evaluate the field from sources lying inside a
sphere. In the method of this paper, we work directly with a spherical
harmonic expansion of the source data. Previous work has used a
Laplace transform method to address the problem of translating data
between concentric
spheres~\cite{field-lau15,greengard-hagstrom-jiang14}. Here we are
interested in transferring data between two spheres with arbitrary
orientation and in the main, the objective is to use source data to
evaluate the spherical harmonic expansion of the incoming acoustic
field on a target sphere.

The established method for performing this operation in the time
domain is the Plane Wave Time Domain
approach\cite{ergin-shanker-michielssen98,ergin-shanker-michielssen99c}
which evaluates the translated field as a superposition of plane
waves. This approach requires careful implementation in order to
remove the anti-causal ``ghost field'' which arises as an artifact of
the plane wave decomposition. A key feature of the method proposed in
this paper is that it is intrinsically causal, by virtue of the
integral approach adopted, and no special treatment is required to
deal with non-physical features.

\section{Algorithm}
\label{sec:algorithm}

\begin{figure}
  \centering
  \includegraphics{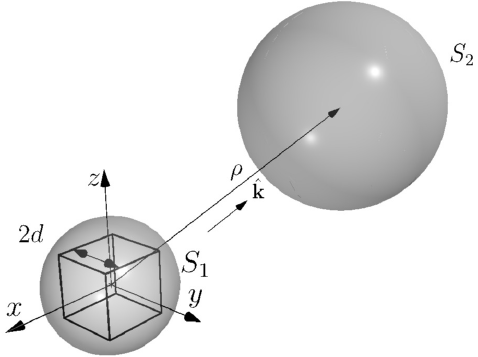}
  \caption{Shift operation between two spherical regions: the acoustic
    field due to sources inside sphere $S_{1}$ is to be evaluated on
    sphere $S_{2}$ which lies at some displacement from
    $S_{1}$. Sources in $S_{1}$ are assumed to lie inside a cube of
    half-width $d$.}
  \label{fig:algorithm:shift}
\end{figure}

The algorithm which we develop in this section is a method for the
estimation of the acoustic field on one sphere $S_{2}$, the
``target'', generated by sources contained inside another
non-intersecting sphere $S_{1}$, the ``source'',
Figure~\ref{fig:algorithm:shift}. As noted above, the principal
motivation for developing the method is its use in a time-domain FMM,
and the notation reflects this motivation, but the method is quite
general and can be used in a range of applications where a field must
be transferred from one origin to another.

The basic elements of the method are representation of surface data
using spherical harmonic expansions and transfer of the field from one
surface to another using the Kirchhoff--Helmholtz integral. Notation
and basic numerical operations are given in the following sections and
then combined to yield an algorithm which is essentially a time-domain
version of the well-known ``point and shoot'' method used in the
single-frequency FMM. 

Figure~\ref{fig:algorithm:shift} shows the configuration and
notation. Sources are contained within a box of half-width $d$,
centered on the origin, which lies inside sphere $S_{1}$. It is
required to determine the field on the second sphere $S_{2}$ whose
center lies at $\rho\khat$. The spheres do not intersect, but one may
lie inside the other. Field data are given as the time-varying
coefficients of a spherical harmonic expansion of acoustic data on the
two spheres: data on $S_{1}$ are determined from the sources inside
the box; data on $S_{2}$ are found using the translation operation,
and can then be used to determine the field inside the sphere. The
steps in the operation are a rotation of the spherical harmonic
expansion on $S_{1}$ to a reference frame aligned with $\khat$;
translation by a distance $\rho$ along the $z$ axis of the rotated
reference frame; rotation of the translated (target) expansion back to
the global frame.

The acoustic field is determined using the Kirchhoff--Helmholtz
integral~\cite[page~182]{pierce89} for the acoustic pressure outside
(or inside) a surface given the acoustic pressure and its normal
derivative on the surface,
\begin{align}
  \label{equ:kirchhoff}
  4\pi p_{2}(\mathbf{x}_{2}, t)
  &=
  \int_{S_{1}} 
  \rhat.\nhat_{1}
  \left(
    \frac{\dot{p}_{1}(\mathbf{x}_{1},\tau)}{Rc}
    +
    \frac{p_{1}(\mathbf{x}_{1},\tau)}{R^{2}}
  \right)
  -
  \frac{1}{R}\frac{\partial p_{1}}{\partial n_{1}}
  \,\D S_{1},\\
  \mathbf{r} &= \mathbf{x}_{2} - \mathbf{x}_{1},
  \,
  R = |\mathbf{r}|,
  \,
  \rhat = \mathbf{r}/R,
  \tau = t - R/c. \nonumber
\end{align}
where $p_{1}$ and $\partial p_{1}/\partial n_{1}$ are the acoustic
pressure and normal derivative on $S_{1}$ and $\nhat_{1}$ is taken as
the outward pointing normal. If the field point $\mathbf{x}_{2}$ is
inside $S_{1}$, the sign of $p_{2}$ is reversed. We also need to
evaluate the normal derivative on $S_{2}$, which can be done by
differentiating \autoref{equ:kirchhoff}:
\begin{align}
  \label{equ:kirchhoff:normal}
  4\pi \frac{\partial p_{2}(\mathbf{x}_{2}, t)}{\partial n_{2}}
  &=
  \int_{S_{1}} 
  \begin{array}[t]{ll}
    \displaystyle
    \left(
      \frac{\dot{p}_{1}}{R^{2}c}
      +
      \frac{p_{1}}{R^{3}}
    \right)
    \nhat_{2}.\nhat_{1} 
    - & \\
    \displaystyle    
    \left(
      \frac{\ddot{p}_{1}}{Rc^{2}} +
      3\frac{\dot{p}_{1}}{R^{2}c} +
      3\frac{p_{1}}{R^{3}}
    \right)
    \rhat.\nhat_{1}\rhat.\nhat_{2}
    + & \\
    \displaystyle
    \left(
      \frac{1}{R^{2}}\frac{\partial p_{1}}{\partial n_{1}}
      +
      \frac{1}{Rc}\frac{\partial \dot{p}_{1}}{\partial n_{1}}      
    \right)
    \rhat.\nhat_{2}
    &
    \,\D S_{1}
  \end{array}
\end{align}
where $\nhat_{2}$ is the normal to surface $S_{2}$ at $\mathbf{x}_{2}$.

The core of the method is an efficient means of translating the
acoustic data along the $z$ axis of the rotated reference frame from
one sphere to another. We now present the notation and basic elements
which form the method.

\subsection{Notation}
\label{sec:notation}

As in previous work~\cite{carley-ghorbaniasl16,carley25} interpolation
of source terms on the enclosing surface is performed using spherical
harmonics, though in the method of this paper operations are applied
directly to the spherical harmonic expansion coefficients, rather than
switching between source representations.

We employ spherical polar coordinates $(\rho,\theta,\phi)$ with
\begin{align*}
  x &= \rho\sin\theta\cos\phi,\, y = \rho\sin\theta\sin\phi,\,
  z = \rho\cos\theta,
\end{align*}
and the spherical harmonics used are
\begin{align*}
  \Ynmi{1}(\theta,\phi) &= \Pbar_{n}^{m}(\theta)\cos m\phi,\\
  \Ynmi{2}(\theta,\phi) &= \Pbar_{n}^{m}(\theta)\sin m\phi,
\end{align*}
with
\begin{align}
  \label{equ:analysis:pbar}
  \Pbar_{n}^{m}(\theta)
  &=
  (-1)^{m}
  \left[
    \frac{2n+1}{4\pi}
    \frac{(n-m)!}{(n+m)!}
  \right]^{1/2}
  P_{n}^{m}(\cos\theta),
\end{align}
where $P_{n}^{m}$ is the associated Legendre function. A function
$f(\theta,\phi)$ defined on the surface of a sphere can then be
represented as
\begin{align*}
  f(\theta,\phi) &=
  \sum_{n=0}^{\infty}\sum_{m=0}^{n}
  \left[
    f_{n,m}^{(1)}\Ynmi{1}(\theta,\phi) +
    f_{n,m}^{(2)}\Ynmi{2}(\theta,\phi)
  \right],\\
  f_{n,m}^{(i)} &=
  \int_{0}^{\pi}\int_{0}^{2\pi}
  f(\theta,\phi)\Ynmi{i}(\theta,\phi)
  \,\D\phi\,\sin\theta\,\D\theta.
\end{align*}

In the numerical implementation, the surface source is evaluated at a
set of quadrature nodes, interpolation coefficients are determined by
a numerical quadrature, and the function is evaluated by a scalar
product with an appropriate vector of spherical harmonic values. We
employ Lebedev quadratures which are near-optimal for integration on
the sphere~\cite{lebedev77,beentjes15}.

We define a spherical harmonic vector $\mathbf{y}(\theta,\phi)$ with
\begin{align}
  y_{2k+i}(\theta,\phi) &= \Ynmi{i}(\theta,\phi),\\
  k &= n(n+1)/2+m, n = 0\ldots,N, m = 0,\ldots,n,\,i=1,2,\\
  f(\theta,\phi) &\approx \mathbf{y}(\theta,\phi).\hat{\mathbf{f}},
\end{align}
where $\hat{\mathbf{f}}$ is the vector of coefficients for the
spherical harmonic expansion evaluated using an appropriate quadrature
rule and $N$ is the order at which the expansion is truncated. Given a
rule for integration on the sphere with nodes $(\theta_{j},\phi_{j})$
and weights $w_{j}$, $j=1,\ldots,N_{Q}$, such that
\begin{align*}
  \int_{0}^{\pi}\int_{0}^{2\pi}
  f(\theta,\phi)
  \,\D S
  &\approx
  \sum_{j=1}^{N_{Q}} f_{j}w_{j},\, f_{j} = f(\theta_{j},\phi_{j}),
\end{align*}
we define a matrix $\mathbf{A}$ with
\begin{align}
  A_{2k+i,j} &= w_{j}\Ynmi{i}(\theta_{j},\phi_{j}),\\
  \hat{\mathbf{f}} &= \mathbf{A}\mathbf{f}.
\end{align}
Given a function sampled at the quadrature nodes on the sphere, the
expansion coefficients can be found by multiplication by $\mathbf{A}$,
and the function can be interpolated if necessary using the vector
$\mathbf{y}(\theta,\phi)$. 

For convenience, we define the matrix $\mathbf{Y}$ for evaluation of
the spherical harmonic expansion at the quadrature nodes,
\begin{align}
  \label{equ:algorithm:Y}
  Y_{j,2k+i} &= \Ynmi{i}(\theta_{j},\phi_{j}),
\end{align}
and a ``rotated'' matrix $\mathbf{Y}^{R}$ which evaluates the
spherical harmonic expansion at transformed nodal positions,
\begin{align}
  \label{equ:algorithm:Y:rotated}
  Y^{R}_{j,2k+i} &= \Ynmi{i}(\theta'_{j},\phi'_{j}),
\end{align}
where $(\theta_{i}',\phi_{i}')$ are coordinates of surface nodes after
application of a rotation operation, to be defined later. This is
equivalent to the method of Lessig et al.\cite{lessig-de-witt-fiume12}
who perform rotation of spherical harmonics by evaluating the data at
rotated sampling locations. A spherical harmonic expansion can then
be rotated using
\begin{align}
  \hat{\mathbf{f}}^{R}
  &=
  \mathbf{R}\hat{\mathbf{f}},\\
  \label{equ:algorithm:rotation}
  \mathbf{R}
  &=
  \mathbf{A}\mathbf{Y}^{R}. 
\end{align}
Other algorithms are available for rotation of spherical harmonics but
for the values of $N$ considered in this paper, this approach has been
found adequate. 

\subsection{Advanced time evaluation}
\label{sec:algorithm:advanced}

The translation method to be described in the following sections uses
discretized impulse responses for circular sources which are evaluated
using an advanced time
algorithm~\cite{casalino03,kessler-wagner04}. In this method, the
source, or retarded, time $\tau$ is taken as the principal variable
and the arrival, or advanced, time $t$ is computed. The signal
generated by an impulse at a single retarded time is then evaluated
over a finite interval of advanced time which depends on the spatial
extent of the source. In this section, we derive the basic elements of
the method for a point source.

\begin{figure}[!htbp]
  \centering
  \includegraphics{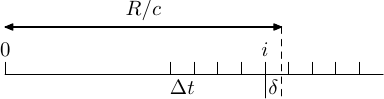}
  \caption{Interpolation of time signal}
  \raggedright
\label{fig:advanced:disc}
\end{figure}

\autoref{fig:advanced:disc} gives notation. A source has time
variation $q(\tau)$ and we evaluate the resulting pressure at some
point as $q(\tau+R/c)/4\pi R$. Time is discretized as $t=i\Delta t$
and from \autoref{fig:advanced:disc}, we set $R/c=(i+\delta)\Delta
t$. For each time step $j$ and corresponding source term $q_{j}$, we
increment the discretized pressure using
\begin{align*}
  p_{i+j+k} \to p_{i+j+k} + \frac{w_{k}}{4\pi R}q_{j},\,k=0,\ldots,K,
\end{align*}
where $i$ accounts for the time delay, $j$ is incremented between time
steps and $w_{k}$ are the weights of a Lagrange interpolation
rule~\cite{berrut-trefethen04,carley25} of order $K$. 

In the remainder of the paper, we write the weight vector as
$\mathbf{w}(R/c)$. We will also need to evaluate the first and second
time derivatives of the pressure, which we do using the Lagrange
differentiation weights~\cite{berrut-trefethen04} with the
corresponding weight vectors denoted $\dot{\mathbf{w}}$ and
$\ddot{\mathbf{w}}$ so that,
\begin{align*}
  \dot{p}_{i+j+k} \to \dot{p}_{i+j+k} + \frac{\dot{w}_{k}}{4\pi R}q_{j},\\
  \ddot{p}_{i+j+k} \to \ddot{p}_{i+j+k} + \frac{\ddot{w}_{k}}{4\pi R}q_{j}.
\end{align*}

\subsection{Transient radiation from a sinusoidal ring source}
\label{sec:algorithm:ring}

\begin{figure}
  \centering
  \includegraphics{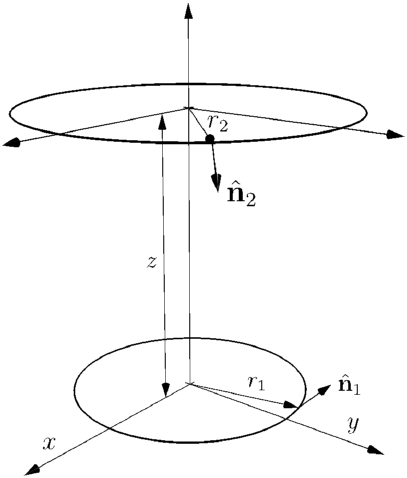}
  \caption{Radiation from source ring (1) to target ring (2)}
  \label{fig:algorithm:ring}
\end{figure}

The translation operation in the next section shifts a spherical
harmonic expansion on a source sphere to a target sphere which lies on
the $z$ axis of the source sphere. In this coaxial coordinate system,
the Kirchhoff--Helmholtz integral reduces to a sequence of
one-dimensional integrals on rings centered on the axis of
translation. On these rings, the source is specified as a Fourier
series in azimuth on the ring and the translation operation yields the
corresponding coefficients of a Fourier series for the resulting field
on a coaxial ring at some axial displacement $z$. 

Figure~\ref{fig:algorithm:ring} defines the notation for radiation
from source ring~1 to target ring~2. Rings have radius $r_{i}$ and
normal $\nhat_{i}$ and field quantities are given by the Fourier series
\begin{align*}
  p_{i}(\phi_{i},\tau)
  &=
  \sum_{m=-\infty}^{\infty}
  \Qmi{i}(\tau)\E^{\J m\phi_{i}},\\
  \frac{\partial p_{i}}{\partial n_{i}}(\phi_{i},\tau)
  &=
  \sum_{m=-\infty}^{\infty}  
  \Nmi{i}(\tau)\E^{\J m\phi_{i}}.
\end{align*}

Each source mode on ring~1 generates a field mode of the same order on
ring~2, with the modal coefficients given by
\begin{align*}
  \Qmi{2}(t)
  &=
  \frac{1}{4\pi}
  \int_{0}^{2\pi}
  \left\{
  \rhat.\nhat_{1}
  \left[
    \frac{\dQmi{1}(\tau)}{Rc} +
    \frac{\Qmi{1}(\tau)}{R^{2}}
  \right]
  -\frac{\Nmi{1}(\tau)}{R}
  \right\}
  \E^{\J m\phi_{1}}
  \,\D\phi_{1},\\
  \Nmi{2}(t)
  &=
  \frac{1}{4\pi}
  \int_{0}^{2\pi}
  \begin{array}[t]{ll}
  \Biggl\{ &
    \displaystyle
    \left[
    \frac{\Qmi{1}(\tau)}{R^{3}} +
    \frac{\dQmi{1}(\tau)}{R^{2}c}
  \right]\nhat_{2}.\nhat_{1} -
  \\
  &
  \displaystyle
  \left[
    3\frac{\Qmi{1}(\tau)}{R^{3}} +
    3\frac{\dQmi{1}(\tau)}{R^{2}c} + 
    \frac{\ddQmi{1}(\tau)}{Rc^{2}}
  \right]\rhat.\nhat_{1}\rhat.\nhat_{2} +
  \\
  &  \displaystyle
  \left[
    \frac{\Nmi{1}(\tau)}{R^{2}} + 
    \frac{\dNmi{1}(\tau)}{Rc}
  \right]\rhat.\nhat_{2}
  \Biggr\}
  \E^{\J m\phi_{1}}
  \,\D\phi_{1},
  \end{array}\\
  \nhat_{i}
  &=
  (n_{ir}\cos\phi_{i}, n_{ir}\sin\phi_{i}, n_{iz}),\,i=1,2,\\
  R^{2} &= r_{1}^{2} + r_{2}^{2} - 2r_{2}r_{1}\cos\phi_{1} + z^{2}.
\end{align*}

Upon discretization and application of the advanced time method of the
previous section, the integrals can be evaluated and at each time step
the modal coefficients on ring~2 are updated using
\begin{subequations}
  \label{equ:algorithm:update}
  \begin{align}
    \Qmi{2}(i+j+k) &\to \Qmi{2}(i+j+k) +
    \cI_{1}^{(m)}(k)\Qmi{1}(j) + \cI_{2}^{(m)}(k)\Nmi{1}(j),\\
    \Nmi{2}(i+j+k) &\to \Nmi{2}(i+j+k) +
    \cI_{3}^{(m)}(k)\Qmi{1}(j) + \cI_{4}^{(m)}(k)\Nmi{1}(j),
  \end{align}  
\end{subequations}
where
\begin{subequations}
  \label{equ:algorithm:ring:int}
  \begin{align}
    \cI_{1}^{(m)}(r_{1},r_{2},z) &=
    \frac{1}{2N_{\phi}}
    \sum_{i=1}^{N_{\phi}}    
    \left[
      \frac{\nhat_{1}.\rhat}{R^{2}}\mathbf{w}(R/c)
      +
      \frac{\nhat_{1}.\rhat}{Rc}\dot{\mathbf{w}}(R/c)
    \right]\cos m\phi_{i},\\
    \cI_{2}^{(m)}(r_{1},r_{2},z) &=
    -\frac{1}{2N_{\phi}}
    \sum_{i=1}^{N_{\phi}}    
    \frac{\mathbf{w}(R/c)}{R}
    \cos m\phi_{i},\\
    \cI_{3}^{(m)}(r_{1},r_{2},z) &=
    \frac{1}{2N_{\phi}}
    \sum_{i=1}^{N_{\phi}}
    \left\{
      (\nhat_{2}.\nhat_{1} - 3\nhat_{1}.\rhat\nhat_{2}.\rhat)
      \left[
        \frac{\mathbf{w}(R/c)}{R^{3}}
        +
        \frac{\dot{\mathbf{w}}(R/c)}{R^{2}c}
      \right]
      -
      \frac{\nhat_{1}.\rhat\nhat_{2}.\rhat}{Rc^{2}}\ddot{\mathbf{w}}(R/c)
    \right\}
    \cos m\phi_{i},\\
    \cI_{4}^{(m)}(r_{1},r_{2},z) &=
    \frac{1}{2N_{\phi}}
    \sum_{i=1}^{N_{\phi}}
    \rhat.\nhat
    \left[
      \frac{\mathbf{w}(R/c)}{R^{2}}
      + 
      \frac{\dot{\mathbf{w}}(R/c)}{Rc}
    \right]
    \cos m\phi_{i}
    \\
    \rhat.\nhat_{1} &=
    \frac{n_{1r}(r_{2}\cos\phi_{i} - r_{1}) + n_{1z}z}{R},\nonumber\\
    \rhat.\nhat_{2} &=
    \frac{n_{2r}(r_{2} - r_{1}\cos\phi_{i}) + n_{2z}z}{R},\nonumber\\
    \nhat_{2}.\nhat_{1}
    &=
    n_{2r}n_{1r}\cos\phi_{i} + n_{2z}n_{1z},\nonumber\\
    \phi_{i} &= 2\pi i/N_{\phi}. \nonumber
\end{align}
\end{subequations}
For any given pair of rings, the vectors $\cI_{i}^{(m)}$ can be
pre-evaluated for $m=0,\ldots,N$ to be used in the integration over
the source sphere developed in the next section. 

\subsection{Axial translation}
\label{sec:algorithm:translation}

\begin{figure}
  \centering
  \includegraphics{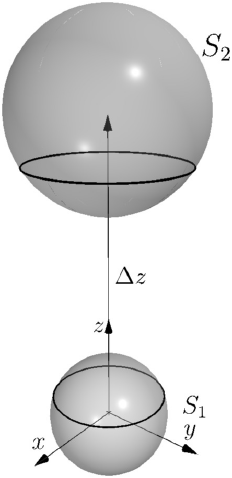}
  \caption{Translation from source to target sphere}
  \label{fig:algorithm:axial}
\end{figure}

After the appropriate rotation, the spherical harmonic coefficients
are translated along the $z$ axis of the coordinate system,
\autoref{fig:algorithm:axial}. This translation is performed by
integrating over the source sphere, which is discretized into ring
sources of the form of \autoref{sec:algorithm:ring}, to evaluate the
Fourier expansion coefficients on rings on the target
sphere. Integration over the target sphere rings then yields the
coefficients of the target spherical harmonic expansion.

Translation is performed using a suitable quadrature rule in $\theta$
on each sphere, $(\theta_{i}^{(1)},w_{i}^{(1)})$ and
$(\theta_{j}^{(2)},w_{j}^{(2)})$, respectively. Fourier coefficients
of the source terms $\Qmi{1}(t)$ and $\Nmi{1}(t)$ are evaluated at
each $\theta_{i}$ using
\begin{align*}
  \Qmi{1}(\theta_{i})
  &=
  \sum_{n=m}^{N}
  \left(
    Q_{n,m}^{(1)} + \J Q_{n,m}^{(2)}
  \right)\Pbar_{n}^{m}(\theta_{i}).
\end{align*}

Target Fourier coefficients at $\theta_{j}$ are then updated using
\begin{subequations}
  \label{equ:algorithm:axial:update}
  \begin{align}
    \Qmi{2}(\theta_{j}) &\to \Qmi{2}(\theta_{j}) +
    \sum_{i}
    w_{i}\sin\theta_{i}
    \left[
      \cI_{1}^{(m)}\Qmi{1}(\theta_{i}) +
      \cI_{2}^{(m)}\Nmi{1}(\theta_{i})
    \right],\\
    \Nmi{2}(\theta_{j}) &\to \Nmi{2}(\theta_{j}) +
    \sum_{i}
    w_{i}\sin\theta_{i}
    \left[
      \cI_{3}^{(m)}\Qmi{1}(\theta_{i}) + \cI_{4}^{(m)}\Nmi{1}(\theta_{i})
    \right].
  \end{align}  
\end{subequations}

Given the target Fourier coefficients, the spherical harmonic
expansion coefficients can be recovered by integration over $\theta$,
\begin{align*}
  Q_{n,m}^{(1)}
  &=
  \real
  \sum_{j}w_{j}\Pbar_{n}^{m}(\theta_{j})\sin\theta_{j}\Qmi{2}(\theta_{j}),\\
  Q_{n,m}^{(2)}
  &=
  \imag
  \sum_{j}w_{j}\Pbar_{n}^{m}(\theta_{j})\sin\theta_{j}\Qmi{2}(\theta_{j}).
\end{align*}

\subsection{Rotation of expansions}
\label{sec:analysis:rotation}

The translation operation described in the previous section shifts the
center of the spherical harmonic expansion along the $z$ axis of the
target sphere. The general translation operation requires a pair of
rotations, one before and one after translation. A number of methods
exist for rotation of spherical harmonic expansions. We employ an
approach similar to that of Lessig et
al.\cite{lessig-de-witt-fiume12}, who rotate expansions by evaluating
the function on rotated sampling points and evaluating the expansion
coefficients. 

We use an auxiliary system of axes $(\ihat,\jhat,\khat)$ with $\khat$
pointing from the source sphere center to the target sphere center,
Figure~\ref{fig:algorithm:shift}. Vectors $\ihat$ and $\jhat$ are
arbitrary unit vectors which form a right-handed triad with $\khat$.

A vector $\mathbf{x}$ is transformed according to
\begin{align*}
  \mathbf{x}' &= (\ihat.\mathbf{x}, \jhat.\mathbf{x}, \khat.\mathbf{x}).
\end{align*}
To evaluate the rotation matrix, each quadrature node $\mathbf{x}_{j}$
on the sphere surface is transformed and its polar coordinates
$(\theta_{j}',\phi_{j}')$ evaluated. These coordinates are then used
in \autoref{equ:algorithm:Y:rotated} and
\autoref{equ:algorithm:rotation} to evaluate the rotation matrix
$\mathbf{R}$, Algorithm~\ref{alg:summary:rotation}. The same approach can be
used to evaluate a rotation matrix for the inverse rotation which is
required as the final step of the method.

\subsection{Summary of algorithm}
\label{sec:analysis:summary}

The elements of the previous sections can now be combined to form the
complete algorithm to shift a spherical harmonic expansion from one
center to another. 

For a given translation distance $\rho$, the operator is precomputed
using Algorithm~\ref{alg:summary:axial}. This gives a set of impulse
responses for each pair of source and target rings in the axial
translation. 

\begin{algorithm}
  \caption{Precompute axial translation operator}
  \label{alg:summary:axial}
  \begin{algorithmic}
    \STATE set $a_{1}$, $a_{2}$, $\rho$, quadrature rules
    $(\theta_{i},w_{i})$ for sphere~1, $(\theta_{j},w_{j})$ for
    sphere~2
    \FOR{$i=1,\ldots,$}
    \STATE $r_{1}=a_{1}\sin\theta_{i}$
    \STATE $z=\rho + a_{2}\cos\theta_{j} - a_{1}\sin\theta_{i}$
    \FOR{$j=1,\ldots,$}
    \STATE $r_{2}=a_{2}\sin\theta_{j}$
    \FOR{$m=0,\ldots,$}
    \STATE evaluate $\cI_{1,2,3,4}^{(m)}(r_{1},r_{2},z)$ and multiply
    by $w_{i}\sin\theta_{i}$ 
    \ENDFOR 
    \ENDFOR    
    \ENDFOR
    \STATE on exit, output is a set of weighted impulse responses for
    each pair of rings on the source and target spheres
  \end{algorithmic}
\end{algorithm}

For a given translation orientation, independent of $\rho$,
Algorithm~\ref{alg:summary:rotation} is used to generate the rotation
matrices for the forward and backward rotations $\mathbf{R}_{F}$ and
$\mathbf{R}_{B}$ respectively.

\begin{algorithm}
  \caption{Precompute rotation operators}
  \label{alg:summary:rotation}
  \begin{algorithmic}
    \STATE input quadrature nodes on sphere, auxiliary coordinate
    system $(\ihat,\jhat,\khat)$, interpolation matrix $\mathbf{A}$
    \FOR{$j=1,\ldots$}
    \STATE set
    $\mathbf{x}_{j}=(\sin\theta_{j}\cos\phi_{j},\sin\theta_{j}\sin\phi_{j},
    \cos\theta_{j})$
    \STATE transform $\mathbf{x}_{j}$ to $(\theta_{j}',\phi_{j}')$ in
    auxiliary coordinate system
    \STATE set row $j$ of
    $\mathbf{Y}^{R}=\mathbf{y}(\theta_{j},\phi_{j})$
    \ENDFOR
    \STATE set $\mathbf{R}=\mathbf{A}\mathbf{Y}^{R}$
    \STATE output is matrix for rotation to coordinate system for
    axial translation
  \end{algorithmic}
\end{algorithm}

To perform the shift operation proper, the spherical harmonic
expansions for the source sphere must be evaluated. Any method can be
used: in this paper we employ Algorithm~\ref{alg:summary:expansion}.

\begin{algorithm}
  \caption{Generation of source spherical harmonic expansions}
  \label{alg:summary:expansion}
  \begin{algorithmic}
    \STATE input source data required to evaluate transient acoustic
    pressure and normal derivative
    \FOR{each time step}
    \FOR{$j=1,\ldots$}
    \STATE evaluate $\phi(\mathbf{x}_{j},t)$, $\partial\phi/\partial
    n(\mathbf{x}_{j},t)$
    \ENDFOR    
    \STATE set $Q_{n,m}(t)=\mathbf{A}\phi(t)$,
    \STATE set $N_{n,m}(t)=\mathbf{A}\partial\phi/\partial n(t)$
    \ENDFOR
    \STATE output is expansion coefficients $\mathbf{Q}=Q_{n,m}$ and
    $\mathbf{N}=N_{n,m}$
    for surface potential and normal derivative respectively at each
    time step
  \end{algorithmic}
\end{algorithm}

Algorithm~\ref{alg:summary:shift} takes the precomputed operators and
the input data and performs the translation. 

\begin{algorithm}
  \caption{Shift operation}
  \label{alg:summary:shift}
  \begin{algorithmic}
    \STATE input source sphere expansion coefficients $\mathbf{Q}$ and
    $\mathbf{N}$ and quadrature rules from
    Algorithm~\ref{alg:summary:axial}
    \STATE apply forward rotation $\mathbf{R}_{F}$ to find source
    expansion coefficients in rotated reference frame
    \FOR{$i=1,\ldots,$}
    \STATE evaluate source azimuthal modes at $\theta_{i}$
    \FOR{$j=1,\ldots,$}
    \FOR{$m=0,\ldots,$}
    \STATE recover $\cI_{1,2,3,4}^{(m)}$ for rings $i$ and $j$ from
    Algorithm~\ref{alg:summary:axial}
    \STATE add contribution of source mode $m$ to target mode $m$ at
    $\theta_{j}$ using Equations~\ref{equ:algorithm:update}
    \ENDFOR    
    \ENDFOR    
    \ENDFOR
    \STATE target azimuthal modes are now evaluated at $\theta_{j}$ on
    $S_{2}$ 
    \FOR{$j=1,\ldots,$}
    \FOR{$n=0,\ldots,N$}
    \FOR{$m=0,\ldots,n$}
    \STATE $Q_{n,m}^{(2)} \to Q_{n,m}^{(2)} + 
    w_{j}\Qmi{2}(\theta_{j})\Pbar_{n}^{m}(\theta_{j})\sin\theta_{j}$
    \STATE $N_{n,m}^{(2)} \to N_{n,m}^{(2)} + 
    w_{j}\Nmi{2}(\theta_{j})\Pbar_{n}^{m}(\theta_{j})\sin\theta_{j}$
    \ENDFOR
    \ENDFOR
    \ENDFOR    
    \STATE apply reverse rotation $\mathbf{R}_{R}$ to find target
    expansion coefficients in global reference frame
  \end{algorithmic}
\end{algorithm}

\subsection{Computational effort}
\label{sec:analysis:time}

The computation time for the translation algorithm can now be
estimated. The two principal elements are the rotation of coefficients
before and after the axial translation step and the axial translation
proper. The rotation is a matrix-vector multiplication for each time
step, or a matrix-matrix multiplication if all time steps are treated
together. The rotation matrices are of size $N(N+1)\times N(N+1)$,
leading to an $O(N^{4})$ computational effort per time step, though
the leading constant is quite small.

The bulk of the computation time is taken by the axial translation
step. The generation of the azimuthal modes on the source sphere takes
$O(N^{2})$ time, as does the transfer of the $N$ source modes to $N$
target modes. This leads to an estimate of $O(N^{2})$ time for the
translation step and for the shift operation as a whole. This is an
effort comparable to established algorithms for harmonic
problems~\cite[chapter~8]{gumerov-duraiswami04}. 

\section{Numerical testing}
\label{sec:results}

We now present results of numerical tests on the translation
method. As noted previously, sources lie inside a cube of half-width
$d$ centered on the origin. This constrains the source sphere radius
to be at least $\sqrt{3}d$, in order to enclose the sources as
required by the Kirchhoff-Helmholtz integral. For testing, we use one
point source positioned on a corner of the cube at $(d,-d,d)$. The
source strength is
\begin{align}
  \label{equ:results:source}
  q(\tau) &= \E^{-\alpha(\tau-\tau_{0})^{2}}\cos\Omega\tau,
\end{align}
with $\Omega=10$, $\alpha=2$, and $\tau_{0}=2$. In all tests, the
acoustic field is evaluated for time $0\leq t\leq 4$. Tests are
performed by evaluating the source spherical harmonic expansion and
translating to a target sphere. A reference calculation is performed
by evaluating the target data directly from the point source and error
is estimated as the difference between the translated and directly
evaluated target distributions. The error measure is the maximum over
all time steps and all values of $n$ and $m$ of the difference between
shifted and directly evaluated coefficients:
\begin{align}
  \label{equ:tests:error}
  \epsilon
  &=
  \frac{\max_{n,m,t} |Q_{2}-Q_{2}'|}{\max_{n,m,t} |Q_{2}'|},
\end{align}
where $Q_{2}$ are the shifted coefficients and $Q_{2}'$ those
evaluated directly from the test source.  Tests were conducted with
fourth order Lagrange interpolation in the advanced time method. Code
was implemented on one core of an Intel i5 using GCC level~2
optimization and OpenBLAS matrix and vector operations.

\begin{figure}
  \centering
  \includegraphics{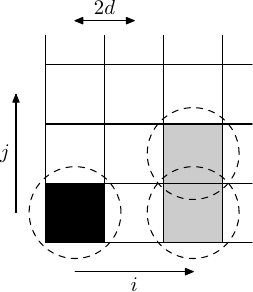}
  \caption{Indexing of boxes for local shift operations: source box is at
    bottom left; target boxes $(2,0,0)$ and $(2,1,0)$ are shown filled
    in gray; enclosing spheres are shown dashed.}
  \label{fig:tests:boxes}
\end{figure}

Test cases are chosen to reflect the principal motivation for
developing the method, the translation operation in the FMM. We assume
that the translation operation is being performed on sources inside
boxes on a regular grid, as shown in~\autoref{fig:tests:boxes}. For a
box of half-width $d$, target boxes are placed at locations denoted
$(i,j,k)$ with centers $(2i,2j,2k)d$. \autoref{fig:tests:boxes} also
indicates the spheres associated with source and target boxes.

An important difference between the integral approach of this paper
and other techniques in the literature is that the acoustic fields
must be evaluated on a sphere surrounding the source or target box,
leading to the requirement to choose a value of the sphere radius. The
lower limit on sphere radius is set by the requirement that the sphere
completely enclose the box; the upper limit is set by the requirement
that source and observer spheres not overlap. In testing the
algorithm, we consider various values of sphere radius within this
range and find that there appears to be a value which optimizes the
accuracy of the translation operation. 


\begin{figure}
  \centering
  \includegraphics{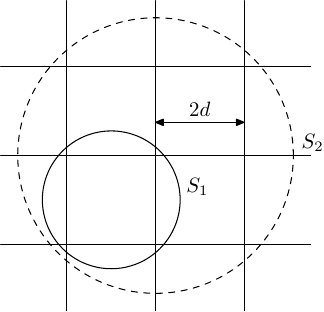}
  \caption{Configuration for upward and downward translation
    operations. For the upward pass, sources lie inside $S_{1}$ and
    the field is translated to $S_{2}$; for the downward pass, sources
    lie outside $S_{2}$ and the field is translated to $S_{1}$.}
  \label{fig:tests:upward}
\end{figure}

The FMM has three main source-target translation operations which must
be evaluated: local, upward, and downward. In a local interaction,
source and target boxes are the same size and quite close ($-3\leq
i,j,k\leq3$), as in \autoref{fig:tests:boxes}. In an upward
interaction, the target sphere is larger than, and encloses the source
sphere; in a downward interaction source and target spheres are
switched, and sources lie outside the source
sphere,~\autoref{fig:tests:upward}. 

Selected results are presented for error in translated expansion
coefficients, varying expansion order $N$, number of time steps
$n_{t}$ and sphere radii $a_{1}$ and $a_{2}$. 

\begin{figure}
  \centering
  \includegraphics{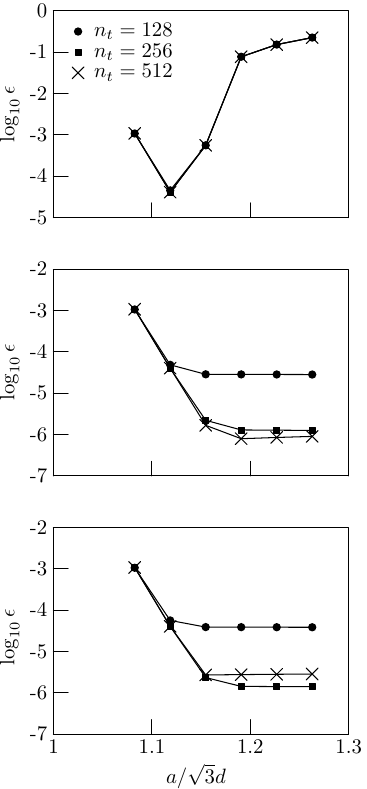}
  \caption{Error in local shift operations against sphere radius with
    $N=40$. Top to bottom: $(2,0,0)$, $(2,1,0)$, $(2,2,2)$ shifts.}
  \label{fig:tests:error:local}
\end{figure}

As noted above, the enclosing spheres must be large enough to enclose
their respective boxes and small enough to avoid overlap. If sources
are contained in a box of half-width $d$, the minimum radius for the
enclosing sphere is $\sqrt{3}d$; the maximum radius depends on the box
separation. The following figures show error as a function of box
radius and how that error varies with box separation and size, as a
function of number of time steps $n_{t}$, with time step $\Delta t =
4/n_{t}$. 

\begin{figure}
  \centering
  \includegraphics{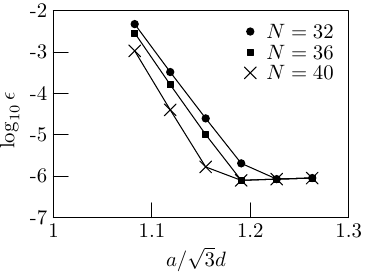}
  \caption{Error in $(2,1,0)$ shift operations against sphere radius
    with varying $N$, $n_{t}=512$.}
  \label{fig:tests:error:N}
\end{figure}

Figures~\ref{fig:tests:error:local} and~\ref{fig:tests:error:N} show
the error in the local shift operation, the translation between two
spheres of the same radius $a$.  We present results for $N=40$ and
three different translations, $(2,0,0)$, $(2,1,0)$, and $(2,2,2)$,
which are representative of typical local interactions in the FMM. The
error for the $(2,0,0)$ case is clearly unacceptable at all values of
sphere radius $a$, and is not reduced by changing $N$ or by reducing
the time step. This is the closest interaction, with one box between
the source and target, see \autoref{fig:tests:boxes}. The conclusion
to be drawn is that this very short-range interaction is not well
computed by the method of this paper and in applications direct
evaluation of the field from the source data would be required.

The other two sets of results, however, show good error behavior with
the error reducing as a function of sphere radius $a$, until it is
constrained by the time discretization. The error is relatively large
at smaller values of $a$. This is because the higher order
coefficients of the spherical harmonic expansion decay more rapidly
for larger values of $a$, i.e. when the source sphere is further from
the point source. Conversely, when $a$ approaches $\sqrt{3}d$, the
higher order coefficients cannot be neglected and at a given value of
$N$, there is a greater error in the source expansion and
correspondingly in the translated target expansion.

Figure~\ref{fig:tests:error:N} presents error in the $(2,1,0)$ shift,
the shortest-range interaction which can be evaluated using the
method, for fixed $n_{t}=512$ and varying expansion order $N$. Again,
the method shows error decreasing until a plateau is reached, with
similar behavior for the three values of $N$ considered. 

\begin{figure}
  \centering
  \includegraphics{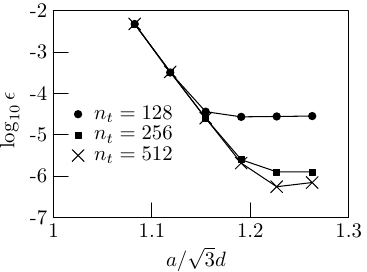}
  \caption{Error in upward shift operation against sphere radius with
    $N=32$.}
  \label{fig:tests:error:upward}
\end{figure}

Figure~\ref{fig:tests:error:upward} shows error in the upward pass
operation where the target sphere encloses the source sphere and is
not concentric with it, Figure~\ref{fig:tests:upward}. In this case,
the target sphere radius is twice the source sphere's,
$a_{2}=2a_{1}$. Again, the error reduces smoothly with $a$ until it is
controlled by the time discretization. 

\begin{figure}
  \centering
  \includegraphics{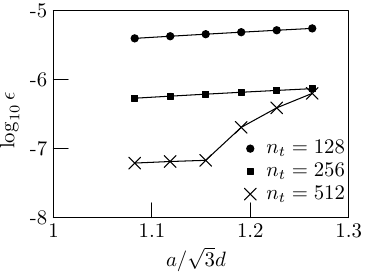}
  \caption{Error in downward shift operation against sphere radius
    with $N=32$.}
  \label{fig:tests:error:downward}
\end{figure}

Figure~\ref{fig:tests:error:downward} shows the error in the downward
pass case. Here, the source sphere encloses the target sphere (the
opposite of the upward pass) and the test source lies outside the
source sphere. The source sphere radius is twice the target
sphere's, $a_{1}=2a_{2}$. In this case, the error eventually increases
with $a$, but remains comparable to that in the downward pass and
local interaction cases up to $a/\sqrt{3}d\approx1.2$, which offers a
hint on the optimal value for sphere radii in applications. 

\begin{figure}
  \centering
  \includegraphics{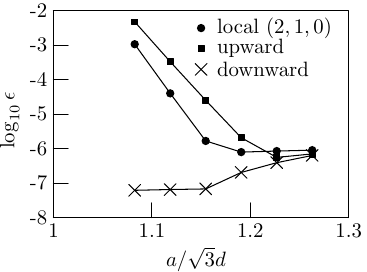}
  \caption{Error against sphere radius for different shifts with
    $N=32$, $n_{t}=512$.}
  \label{fig:tests:compare}
\end{figure}

Figure~\ref{fig:tests:compare} compares the error at fixed $N$ and
$n_{t}$ for the three translations considered. There is a region from
$1.15\lessapprox a/\sqrt{3}d\lessapprox 1.25$ where the error for all
three shifts is comparable and less than about $10^{-6}$. This
indicates that there is a sphere enclosing the source box whose radius
is large enough to limit the error in evaluating the source terms
without being so close to the target sphere that the accuracy of the
transfer operation is affected.

\begin{figure}
  \centering
  \includegraphics{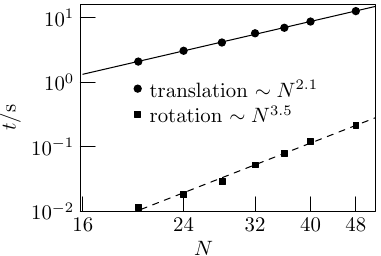}
  \caption{Computation time for translation and rotation
    operations against spherical harmonic expansion order.}
  \label{fig:tests:timing}
\end{figure}

\autoref{fig:tests:timing} shows the computation time for the rotation
and translation stages of the shift operation as a function of
expansion order $N$ with $n_{t}=512$ time steps. Computation times are
estimated as the average over all values of source and target radius
for the $(2,1,0)$ shift whose results are presented in
\autoref{fig:tests:error:N}. Fitted data show that time for the
rotation operation scales as $N^{3.5}$ but with a much smaller leading
constant than the axial translation whose computation time scales as
$N^{2.1}$. Overall calculation time thus scales approximately as
$N^{2}$, consistent with the estimates of \autoref{sec:analysis:time},
and making the algorithm comparable to existing translation methods
for the time-harmonic and Laplace problems.

\begin{figure}
  \centering
  \includegraphics{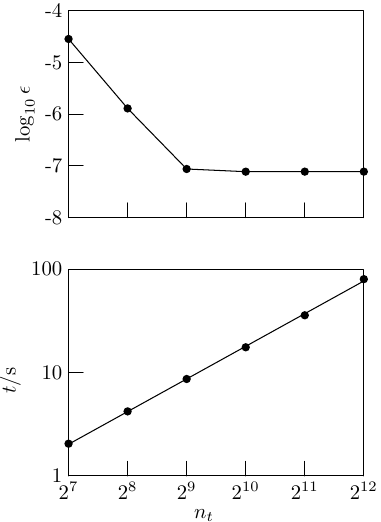}
  \caption{Error (upper plot) and total computation time (lower plot)
    against number of time steps for $(2,1,0)$ local shift with
    $N=40$; lower plot shows linear fit to computation time.}
  \label{fig:tests:stability}
\end{figure}

Finally, \autoref{fig:tests:stability} presents results for the
performance of the method as a function of the number of time steps,
$128\leq n_{t}\leq 4096$ ($2^{-10}\leq\Delta t\leq 2^{-5}$). The lower
plot, of computation time, shows the linear scaling which would be
expected from an algorithm of this type: the computational effort per
source time point is constant and so the total time scales linearly on
the number of time steps. The upper plot showing error as a function
of $n_{t}$ shows the same behavior as in
Figures~\ref{fig:tests:error:local} and~\ref{fig:tests:error:upward},
with error reducing with time step until it reaches a minimum. More
importantly, it demonstrates that the method is stable as the time
step is reduced, and does not require the specialized approaches
needed to avoid instability in convolution-based methods.

\section{Conclusions}
\label{sec:conclusions}

A stable causal method has been developed for the translation of
acoustic field data from a source to a target region, using the
Kirchhoff--Helmholtz integral applied to a spherical harmonic
expansion of surface data on spheres surrounding the source and target
regions respectively. The motivation for the method is the shift
operation in a time-domain FMM, and test results reflecting this
motivation have been presented. For the expansion orders $N$
considered, the computation time scales approximately as $N^{2}$. The
approach does not work well for the very closest interactions, but we
otherwise find that it converges with six digit accuracy for suitable
values of the parameters of the method.

\section{Acknowledgments}
\label{sec:acknowledgments}

The author is indebted to the anonymous reviewer whose careful reading
of the manuscript has improved the clarity of the presentation and has
helped highlight a number of points of interest.

\section{Author declarations}
\label{sec:declarations}

The author has no conflicts of interest to declare.

\section{Data availability}
\label{sec:data:availability}

Code implementing the method of the paper and generating the results
presented is available upon request to the author.

\end{document}